\documentclass[review,3p]{elsarticle}
\usepackage{amsfonts}
\usepackage{amsfonts,amssymb,mathrsfs,amsmath}
\usepackage[colorlinks]{hyperref}
\usepackage{lineno,graphicx,epstopdf}

\newtheorem{thm}{Theorem}
\newtheorem{lem}[thm]{Lemma}

\newtheorem{pro}[thm]{Proposition}
\newdefinition{rmk}{Remark}
\newdefinition{exam}{Example}
\newdefinition{df}{Definition}
\newproof{pf}{\noindent{\bf Proof}}
\def\qed{\quad$\Box$}

\usepackage{amssymb}

\journal{}

\begin{document}

\begin{frontmatter}

\title{{\bf Simple characterizations for commutativity of quantum weakest preconditions}}

\author[a]{Tianrong Lin\corref{cor1}}

\address[a]{Fukien University of Technology GM Information College, Fukien, China}

\begin{abstract}
In a recent letter [Information Processing Letters~104 (2007) 152-158], it has shown some sufficient conditions for commutativity of quantum weakest preconditions. This paper provides some alternative and simple characterizations for the commutativity of quantum weakest preconditions, i.e., Theorem \ref{thm3}, Theorem \ref{thm4} and Proposition \ref{pro5} in what follows. We also show that to characterize the commutativity of quantum weakest preconditions in terms of $[M,N]$ ($=MN-NM$) is hard in the sense of Proposition \ref{pro7} and Proposition \ref{pro8}.
\end{abstract}


\end{frontmatter}


\section{Introduction}\label{sec1}

 \begin{figure}[h!]
	     \centering
	     \includegraphics[width=0.93\textwidth]{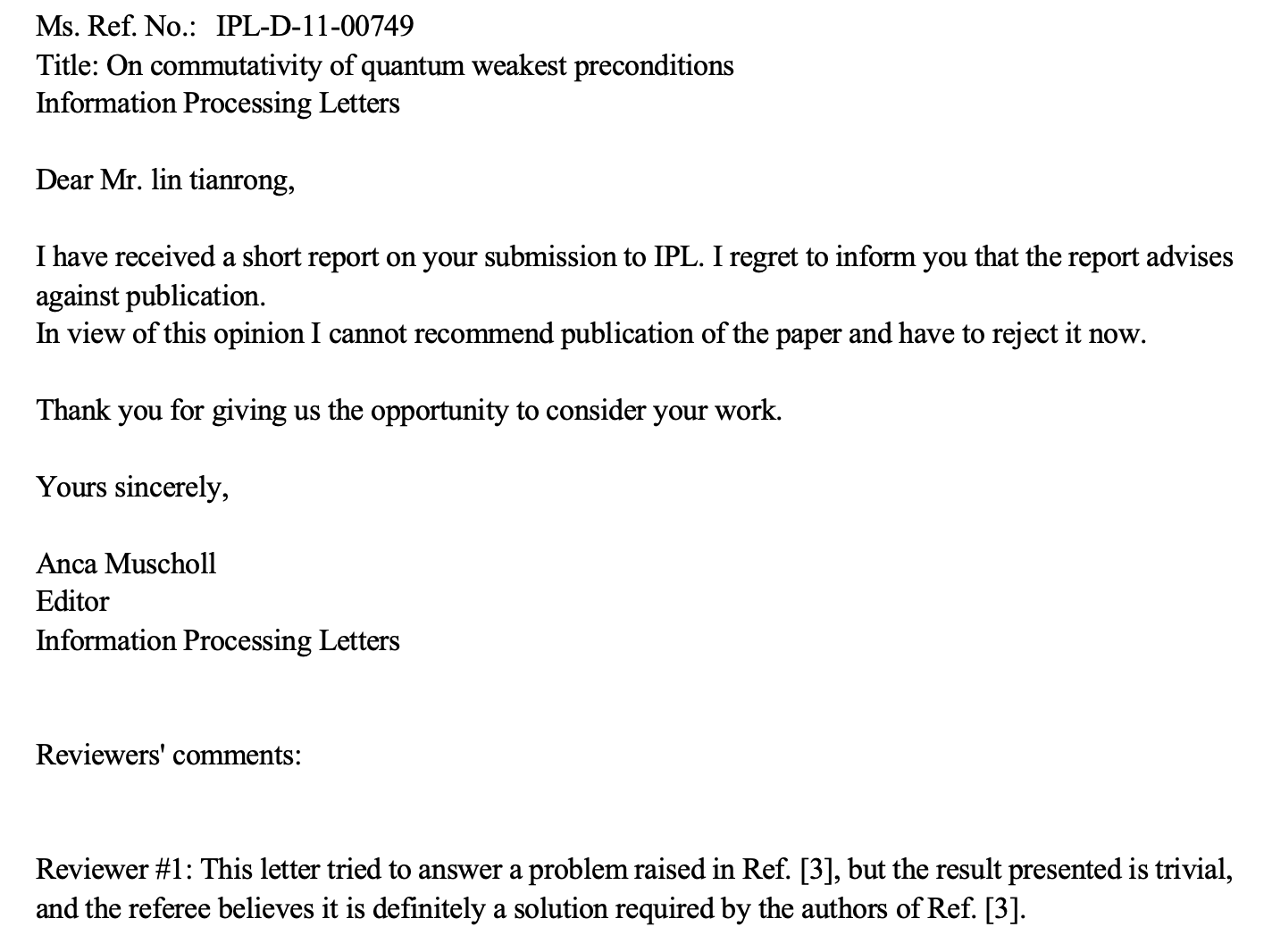}
	     \caption{\label{fig:Review Report in 2012} Review Report from the authors of Ref. \cite{3}}
       \end{figure}

The theory of quantum computation, including the subfield of semantics for quantum programming languages \cite{1,2,3,4,5,6,7}, develops rapidly. This is, to a large extent, owing to the motivation of Shor's quantum factoring algorithms \cite{11} and Grover's searching algorithm \cite{12}.

\par
Quantum algorithms is a very important research direction \cite{13}. However, quantum algorithm currently are expressed at the very low level of quantum circuits \cite{13} which is a disadvantage in some research situations. To make progress, scientists have contributed their enormous efforts to investigate design and semantics of quantum programming languages \cite{1,2,3,4,5,6,7}, so that quantum algorithm can be represented at relatively hight level of quantum programming languages.

\par
In Ref.~\cite{4}, D'Hondt and Panangaden introduced a notion of quantum weakest precondition and a Stone-type duality between the state transition semantics and the predicate transformer semantics for quantum programs. In their approaches, a quantum predicate is defined to be an observable, i.e., a Hermitian operator on the state space, which can be seen as a natural generalization of Kozen's probabilistic predicate as a measurable function \cite{14}. According to Selinger's viewpoint \cite{3}, quantum programs may be represented by super-operators. Then, D'Hondt and Panangaden showed that quantum weakest precondition can be expressed in terms of operators of quantum programs (i.e., super-operators) and a fixed Hermitian operator \cite{3}.

\par
Our main attention in this paper is the commutativity of quantum weakest preconditions. As the observation of Ying et al.'s \cite{5,6} claimed, quantum predicate transformer semantics is not a simple generalization of predicate transformer semantics for classical and probabilistic programs and we should to answer some important problems that would not arise in the realm of classical and probabilistic programming. Of such problems that are known to be important, the commutativity of quantum weakest preconditions is urgent to be answered, since just as mentioned in \cite{6}, the physical simultaneous verifiability of quantum weakest preconditions depends on commutativity between them according to the Heisenberg uncertainty principle.

\par
This paper provides three simple characterizations for commutativity of quantum weakest preconditions. The main idea is that we should characterize the commutativity of quantum weakest preconditions in terms of the properties of quantum weakest preconditions rather than the $[M,N]$. The most obvious property of quantum weakest precondition is that quantum weakest precondition is again an observable, (see Lemma \ref{lem6} in the sequel), i.e. a Hermitian operator on the state space, although we often forgot this fact in practice.

\par
Let $wp(\mathcal{E})(M)$ and $wp(\mathcal{E}(N))$ be two quantum weakest preconditions. Then, this paper will show in Section \ref{sec3} the following

\begin{itemize}
  \item {$wp(\mathcal{E})(M)$ and $wp(\mathcal{E})(N)$ commute if and only if the product of them is Hermitian;}
  \item {$wp(\mathcal{E})(M)$ and $wp(\mathcal{E})(N)$ commute if and only if there exists an Unitary matrix $U$ such that \[U^{\dagger}wp(\mathcal{E})(M)U=\mbox{diag}(\lambda_1,\cdots,\lambda_n)\] and \[U^{\dagger}wp(\mathcal{E})(N)U=\mbox{diag}(\mu_1,\cdots,\mu_n),\] where $\lambda_i$ and $\mu_i$ are the eigenvalues of $wp(\mathcal{E})(M)$ and $wp(\mathcal{E})(N)$, respectively; }
  \item {$wp(\mathcal{E})(M)$ and $wp(\mathcal{E})(N)$ commute if and only if  
   \[\mbox{Tr} \footnote{\[\mbox{Tr}(A)=\sum\limits_{i=1}^na_{ii}\] where $A=(a_{ij})_{n\times n}$.}\left(\Big(wp(\mathcal{E})(M)wp(\mathcal{E})(N)\Big)^2\right)=\mbox{Tr}\left(\Big(wp(\mathcal{E})(M)\Big)^2\Big(wp(\mathcal{E})(N)\Big)^2\right),\] where $wp(\mathcal{E})(M)$ is a quantum weakest precondition.}
\end{itemize}

\par
We would like to point out that the above results seems to be trivial (thus simple). Indeed, (2) and (3) come naturally from some facts of linear algebra \cite{15}, so long as the reader recalls that a quantum weakest precondition is again a Hermitian matrix. One may naturally expect that whether we can characterize the commutativity of quantum weakest preconditions in terms of $[M,N]$ (=$MN-NM$), because a quantum weakest precondition is represented by a Hermitian matrix and operators of a super-operator (see, Proposition \ref{pro1}) (Perhaps there are some other reasons). However, the examples illustrated in this paper show that this may be very difficult (see, Proposition \ref{pro7} and Proposition \ref{pro8}).

\par
The remainder of the paper is organized in the following way: the next Section is devoted to review some basic definitions and useful propositions where the main results are introduced. Section \ref{sec3} is devoted to the proofs of the main results where some examples are presented, and Section \ref{sec4} is the concluding Section.

\section{Preliminaries and main results}\label{sec2}

Let $\mathcal{H}$ be a Hilbert space. Recall in \cite{6} that a density matrix $\rho$ on $\mathcal{H}$ is a positive operator with Tr$(\rho)\leq 1$.  The set of density operators on $\mathcal{H}$ is denoted $\mathcal{D}(\mathcal{H})$. A super-operator on a Hilbert space $\mathcal{H}$ is a linear operator $\mathcal{E}$ from the space $\mathcal{L}(\mathcal{H})$ into itself which the following are satisfied
\begin{itemize}
  \item {Tr[$\mathcal{E}(\rho)$]$\leq$Tr($\rho$) for each $\rho\in\mathcal{D}(\mathcal{H})$;}
  \item {Complete positivity: for any extra Hilbert space $\mathcal{H}_R$, $(\mathcal{I}_R\otimes \mathcal{E})(A)$ is positive provided $A$ is a positive operator on $\mathcal{H}_R\otimes\mathcal{H}$, where $\mathcal{I}_R$ is the identity operation on $\mathcal{H}_R$.}
\end{itemize}

Analogue to \cite{6}, the set of super-operators on $\mathcal{H}$ is denoted as $\mathcal{C}\mathcal{P}(\mathcal{H})$.

\par
A quantum predicate on $\mathcal{H}$ is defined to be a Hermitian operator $M$ with
$0\sqsubseteq M\sqsubseteq I$, where the ordering ``$\sqsubseteq$" is the L\"owner ordering, i.e., $A\sqsubseteq B$ if $B-A$ is a positive operator. The set of quantum predicates on $\mathcal{H}$ is denoted by $\mathcal{P}(\mathcal{H})$.

\par
We state the definition of precondition of a quantum predicate $N$ with respect to a quantum program $\mathcal{E}$ in the following

\begin{df}[\cite{6}, Definition 2.1]\label{df1}
 For any quantum predicates $M,N\in\mathcal{P}(\mathcal{H})$, and for any quantum program $\mathcal{E}\in\mathcal{C}\mathcal{P}(\mathcal{H})$, $M$ is called a precondition of $N$ with respect to $\mathcal{E}$, written $M\{\mathcal{E}\}N$, if
 \[\mathrm{Tr}(M\rho)\leq\mathrm{Tr}(N\mathcal{E}(\rho))\]
 for all density operator $\rho\in\mathcal{D}(\mathcal{H})$.
\end{df}

\par
With the above Definition \ref{df1} in mind, we introduce the concept of quantum weakest precondition as follows.

\begin{df}[\cite{6}, Definition 2.2]\label{df2}
Let $M\in\mathcal{P}(\mathcal{H})$ be a quantum predicate and $\mathcal{E}
\in\mathcal{C}\mathcal{P}(\mathcal{H})$ a quantum program. Then the weakest precondition of $M$ with respect to $\mathcal{E}$ is a quantum predicate $wp(\mathcal{E})(M)$ satisfying the following conditions:
\begin{enumerate}
  \item {$wp(\mathcal{E})(M)\{\mathcal{E}\}M$;}
  \item {for all quantum predicates $N$, $N\{\mathcal{E}\}M$ implies $N\sqsubseteq wp(\mathcal{E})(M)$.}
\end{enumerate}
\end{df}

The following Kraus operator-sum representation of $wp(\mathcal{E})$ is necessary in the sequel.

\begin{pro}[\cite{6}, Proposition 2.1]\label{pro1}
  Suppose that $\mathcal{E}\in\mathcal{C}\mathcal{P}(\mathcal{H})$ is represented by the set $\{E_i\}$ of operators. Then for each $M\in\mathcal{P}(\mathcal{H})$, we have
  \begin{equation}
  \label{wqp}
  \begin{split}
    wp(\mathcal{E})(M)=&\sum_iE_iME_i^{\dagger}.
  \end{split}
  \end{equation}
\end{pro}

\par
The following intrinsic characterization of $wp(\mathcal{E})$, attributed to Ying et al. \cite{6} (also, \cite{5}), deals with the case that $\mathcal{E}$ is given by a system-environment model\footnote[5]{Cf. \cite{6}, {\bf Theorem 2.1} item 2.}.

\begin{pro}[\cite{6}, Proposition 2.2]\label{pro2}
  If $\mathcal{E}$ is given in terms of system-environment model, then we have
   \[ wp(\mathcal{E})(M)=\langle e_0|U^{\dagger}P(M\otimes I_E)PU|e_0\rangle\]
  for each $M\in\mathcal{P}(\mathcal{H})$, where $I_E$ is the identity operator in the environment system.
\end{pro}

\par
Now, we turn to the notion of {\it commutativity} for two operators. In general, two operators $A$ and $B$ on $\mathcal{H}$ are said to be commutative if $AB=BA$, i.e., $[A,B]=AB-BA=0$. Restrict our attention to quantum weakest preconditions, we said that two quantum weakest preconditions $wp(\mathcal{E})(M)$ and $wp(\mathcal{E})(N)$ commutative if $wp(\mathcal{E})(M)wp(\mathcal{E})(N)=wp(\mathcal{E})(N)wp(\mathcal{E})(M)$. The issue we want to dealt with in the present paper is that under what conditions two quantum weakest preconditions $wp(\mathcal{E})(M)$ and $wp(\mathcal{E})(N)$ commute.

\par
The following are the main results of this paper.

\begin{thm}\label{thm3}
  Let $M,\,N\in\mathcal{P}(\mathcal{H})$ be two quantum predicates, and  $\mathcal{E}\in\mathcal{C}\mathcal{P}(\mathcal{H})$ a quantum program. Then $wp(\mathcal{E})(M)$ and $wp(\mathcal{E})(N)$ commute iff the product $wp(\mathcal{E})(M)\cdot wp(\mathcal{E})(N)$ (or, $wp(\mathcal{E})(N)\cdot wp(\mathcal{E})(M)$) is Hermitian.
\end{thm}

\begin{thm}\label{thm4}
  Let $M,\,N\in\mathcal{P}(\mathcal{H})$ be two quantum predicates, and  $\mathcal{E}\in\mathcal{C}\mathcal{P}(\mathcal{H})$ a quantum program. Then $wp(\mathcal{E})(M)$ and $wp(\mathcal{E})(N)$ commute iff there exists an Unitary matrix $U$ such that \[U^{\dagger}wp(\mathcal{E})(M)U=\mathrm{diag}(\lambda_1,\cdots,\lambda_n)\] and \[U^{\dagger}wp(\mathcal{E})(N)U=\mathrm{diag}(\mu_1,\cdots,\mu_n)\] where $\lambda_i$ and $\mu_i$ are the eigenvalues of $wp(\mathcal{E})(M)$ and $wp(\mathcal{E})(N)$, respectively.
\end{thm}

\par
The following Proposition borrowed from \cite{15} is an another characterization for commutativity of $wp(\mathcal{E})(M)$ and $wp(\mathcal{E})(N)$ (cf.~\cite{15}, p.~552, exerc.~6).

\begin{pro}\label{pro5}
 Let $M,\,N\in\mathcal{P}(\mathcal{H})$ be two quantum predicates, and  $\mathcal{E}\in\mathcal{C}\mathcal{P}(\mathcal{H})$ a quantum program. Then $wp(\mathcal{E})(M)$ and $wp(\mathcal{E})(N)$ commute iff \[\mathrm{Tr}\Big((wp(\mathcal{E})(M)wp(\mathcal{E})(N))^2\Big)=\mathrm{Tr}\Big((wp(\mathcal{E})(M))^2(wp(\mathcal{E})(N))^2\Big).\]
\end{pro}

\section{Proofs of the main results} \label{sec3}

As mentioned in Sect. \ref{sec1}, quantum weakest preconditions are again observations. This can be seen from the following lemma which also plays a crucial role in the proofs of Theorem \ref{thm3}, Theorem \ref{thm4} and Proposition \ref{pro5}.

\begin{lem}\label{lem6}
  Let $M\in\mathcal{P}(\mathcal{H})$ be a quantum predicate, and let $\mathcal{E}\in\mathcal{C}\mathcal{P}(\mathcal{H})$ be a quantum program. Then $wp(\mathcal{E})(M)$ is Hermitian.
\end{lem}

\begin{pf}
Let $A$ denote $wp(\mathcal{E})(M)$, then by Proposition \ref{pro1},
\begin{equation}
\begin{split}
  A^{\dagger}=&\left(\sum_iE_iME_i^{\dagger}\right)^{\dagger}=\sum_i\Big(E_iME_i^{\dagger}\Big)^{\dagger}\nonumber\\
             =&\sum_i(E_i^{\dagger})^{\dagger}M^{\dagger}E_i^{\dagger}=\sum_iE_iME_i^{\dagger}\\
             =&A
    \end{split}
  \end{equation}
as required.\qed
\end{pf}

\par
Now, we can present the proofs of main results as follows.\\

\par
\noindent{\bf Proof of Theorem \ref{thm3}.} Let $A$ denote $wp(\mathcal{E})(M)$ and $B$ denote $wp(\mathcal{E})(N)$, respectively.

  \par
  We show first the ``if" part of the Theorem.

  \par
  If $AB$ is Hermitian, i.e., $(AB)^{\dagger}=AB$, then
 \begin{equation}
  \begin{split}
    AB=&(AB)^{\dagger}=B^{\dagger}A^{\dagger}\nonumber\\
      =&BA\qquad\mbox{(by {\bf Lemma \ref{lem6}})}
    \end{split}
  \end{equation}

  \par
  We show next the ``only if" part of the Theorem.
  \par
  Assume that $A$ and $B$ commute, i.e., $AB=BA$, then
  \begin{equation}
  \begin{split}
    (AB)^{\dagger}=&B^{\dagger}A^{\dagger}
                  =BA\qquad\mbox{(by {\bf Lemma \ref{lem6}})}\nonumber\\
                  =&AB\qquad\mbox{(by Hypothesis)}
    \end{split}
  \end{equation}
 Theorem \ref{thm3} follows.\qed\\

\par
\noindent{\bf Proof of Theorem \ref{thm4}.} Also, let $A$ denote $wp(\mathcal{E})(M)$ and $B$ denote $wp(\mathcal{E})(N)$, respectively.
\par
The ``if" part is obvious.

\par
We show the ``only if" part. Viewed $A$, $B$ as linear transformations on the vector space $\mathbb{C}^n$. Let $\mathcal{V}=\{A\alpha:\,\,\alpha\in\mathbb{C}^n\}$ and assume that $AB=BA$.  Then for any $\beta\in\mathcal{V}$, we see that
\begin{equation}
\begin{split}
B\beta=&B\big(A\alpha\big)=AB\alpha\quad\mbox{(by hypothesis)}\nonumber\\
=&A(B\alpha)\in\mathcal{V}
\end{split}
\end{equation}

That is, $\mathcal{V}$ is an invariant subspace under transformation $B$. We further assume that dim$\mathcal{V}=n_1\leq n$. Then, there exist $n_1$ unit vectors $\{|e_1\rangle,\cdots,|e_{n_1}\rangle\}\subset\mathcal{V}$ such that
  \[B|e_i\rangle=\mu_i|e_i\rangle\quad\mbox{for all $1\leq i\leq n_1$},\]
because $\mathcal{V}$ can be decomposed to $\mathcal{V}=V_{\lambda_1}\oplus\cdots\oplus V_{\lambda_{n_1}}$ where $V_{\lambda_i}=\{\zeta:\,\,A\zeta=\lambda_i\zeta\}$. It is easy to see that for any $\zeta\in V_{\lambda_i}$, $B\zeta\in V_{\lambda_i}$, i.e., $V_{\lambda_i}$ is an invariant subspace under transformation $B$.\footnote{It is clear from the fact that: $A(B\zeta)=B(A\zeta)=B(\lambda_i\zeta)=\lambda_i(B\zeta)$}

\noindent Further, let $\{|e_{n_1+1}\rangle,\cdots,|e_n\rangle\}$ be a basis of vector space $\mathcal{V}^{\bot}$ \footnote{$\mathcal{V}^{\bot}$ is defined as $\{|\alpha\rangle:\,|\alpha\rangle\in\mathbb{C}^n,\,\langle\beta|\alpha\rangle=0$ for any $|\beta\rangle\in\mathcal{V}\}$.} such that
\begin{itemize}
  \item {$B|e_j\rangle=\mu_j|e_j\rangle$ for all $n_1+1\leq j\leq n$;}
  \item {$
\{|e_1\rangle,\cdots,|e_{n_1}\rangle,|e_{n_1+1}\rangle,\cdots,|e_n\rangle\}$
is an orthogonal basis of $\mathbb{C}^n$.}
\end{itemize}

\par
 Let
\[U=\left(|e_1\rangle,\cdots,|e_{n_1}\rangle,|e_{n_1+1}\rangle,\cdots,|e_n\rangle\right)^{\dagger}\]
Then it is easy to derive the following
\[UBU^{\dagger}=\mbox{diag}(\mu_1,\cdots,\mu_{n_1},\mu_{n_1+1}\cdots,\mu_n)\]
and
\[UAU^{\dagger}=\mbox{diag}(\lambda_1,\cdots,\lambda_{n_1},0,\cdots,0)\]

\noindent Theorem \ref{thm4} follows.\qed\\

\par
\noindent{{\bf Proof of Proposition \ref{pro5}.}} Let $A$ denote $wp(\mathcal{E})(M)$ and $B$ denote $wp(\mathcal{E})(N)$, respectively. Then, the ``only if" part is obvious.

 \par
 To show the ``if" part of proposition, note first that Tr$(MM^{\dagger})=0$ if and only if $M=0$ where $M$ is a square matrix of size $n$ over $\mathbb{C}$. Then, by the assumption that Tr$((AB)^2)=\mathrm{Tr}(A^2B^2)$ and Lemma \ref{lem6}, we can easy see the following
  \[\mathrm{Tr}\Big((AB-BA)(AB-BA)^{\dagger}\Big)=0\]
 Hence, $AB-BA=0$, i.e., $AB=BA$.\qed

\par
\begin{rmk}
Since the weakest quantum preconditions are given in the form of Eq.~(\ref{wqp}), a natural idea to characterize the commutativity of them is in terms of $[M,N]$, i.e., the commutator of $M,N$. Unfortunately, the following examples show that this is difficult.
\end{rmk}
\par
\begin{exam}\label{exam1}
 Let
\begin{eqnarray*}
 M=\left(
     \begin{array}{cc}
       .2 & .2i \\
       -.2i & .5 \\
     \end{array}
   \right)\qquad
 N=\left(
     \begin{array}{cc}
       .3 & .1+.2i \\
       .1-.2i & 0 \\
     \end{array}
   \right)\qquad
   \mathcal{E}=\left\{\left(
        \begin{array}{cc}
          .1 & 0 \\
          0 & 0 \\
        \end{array}
      \right)\right\}
\end{eqnarray*}
where $i=\sqrt{-1}$. Then, it is easy to verify that
\begin{equation}
\begin{split}
 MN=&\left(
     \begin{array}{cc}
       .2 & .2i \\
       -.2i & .5 \\
     \end{array}
   \right)\left(
     \begin{array}{cc}
       .3 & .1+.2i \\
       .1-.2i & 0 \\
     \end{array}
   \right)=\left(
                \begin{array}{cc}
                  .1+.02i & .02+.04i \\
                  .05-.16i & .04-.02i \\
                \end{array}
              \right)
     \nonumber \\
\neq& \left(
                \begin{array}{cc}
                  .1-.02i & .05+.16i \\
                  .02-.04i & .04+.02i \\
                \end{array}
              \right)=\left(
     \begin{array}{cc}
       .3 & .1+.2i \\
       .1-.2i & 0 \\
     \end{array}
   \right)\left(
     \begin{array}{cc}
       .2 & .2i \\
       -.2i & .5 \\
     \end{array}
   \right)=NM.
   \end{split}
\end{equation}
However,
\begin{equation}
\begin{split}
wp(\mathcal{E})(M)=&\left(
  \begin{array}{cc}
    .1 & 0 \\
    0 & 0 \\
  \end{array}
\right)\left(
                   \begin{array}{cc}
                     .2 & .2i \\
                     -.2i & .5 \\
                   \end{array}
                 \right)\left(
                          \begin{array}{cc}
                            .1 & 0 \\
                            0 & 0 \\
                          \end{array}
                        \right)^{\dagger}
                                       =\left(
                                                 \begin{array}{cc}
                                                   .002 & 0 \\
                                                   0 & 0 \\
                                                 \end{array}
                                               \right)\nonumber\\
wp(\mathcal{E})(N)=&\left(
  \begin{array}{cc}
    .1 & 0 \\
    0 & 0 \\
  \end{array}
\right)\left(
     \begin{array}{cc}
       .3 & .1+.2i \\
      .1-.2i & 0 \\
     \end{array}
   \right)\left(
                          \begin{array}{cc}
                            .1 & 0 \\
                            0 & 0 \\
                          \end{array}
                        \right)^{\dagger}
                                       =\left(
                                                 \begin{array}{cc}
                                                 .003  & 0 \\
                                                 0 & 0 \\
                                                 \end{array}
                                               \right)
\end{split}
\end{equation}
which means that
 \[wp(\mathcal{E})(M)\cdot wp(\mathcal{E})(N)=wp(\mathcal{E})(N)\cdot wp(\mathcal{E})(M).\]
\end{exam}

\par
The above Example \ref{exam1} implies the following

\begin{pro}\label{pro7}
 For any $n\geq 2$, there exist $M,N\in\mathcal{P}(\mathcal{H}_n)$ with $\mathrm{ dim}\mathcal{H}_n$=n, and $\mathcal{E}\in\mathcal{C}\mathcal{P}(\mathcal{H}_n)$ such that
  \[wp(\mathcal{E})(M)\cdot wp(\mathcal{E})(N)=wp(\mathcal{E})(N)\cdot wp(\mathcal{E})(M)\]
 with $MN\neq NM$.
\end{pro}
\begin{pf}
  Let
  \begin{equation}
  \begin{split}
    M=\left(
        \begin{array}{cc}
          I_{n-2} & 0 \\
          0 & \left(
                \begin{array}{cc}
                  .2 & .2i \\
                  -.2i & .5 \\
                \end{array}
              \right)
           \\
        \end{array}
      \right)\qquad N=\left(
                        \begin{array}{cc}
                          I_{n-2} & 0 \\
                          0 & \left(
                                \begin{array}{cc}
                                  .3 & .1+.2i \\
                                  .1-.2i & 0 \\
                                \end{array}
                              \right)
                           \\
                        \end{array}
                      \right)\nonumber
      \end{split}
  \end{equation}
  and
  \begin{equation}
  \begin{split}
  \mathcal{E}=\left\{\left(
                         \begin{array}{cc}
                           I_{n-2} & 0 \\
                           0 & \left(
                                  \begin{array}{cc}
                                    .1 & 0 \\
                                    0 & 0 \\
                                  \end{array}
                                \right)
                            \\
                         \end{array}
                       \right)\right\}\nonumber
  \end{split}
  \end{equation}
  where $I_{n-2}$ is the unit matrix of size $n-2$. \qed
\end{pf}

\par
\begin{exam}\label{exam2}
 Let
\begin{equation}
\begin{split}
  M=\left(
      \begin{array}{cc}
        .2 & 0 \\
        0 & 0 \\
      \end{array}
    \right)\qquad
  N=\left(
      \begin{array}{cc}
        .3 & 0 \\
        0 & .7 \\
      \end{array}
    \right)\qquad
    \mathcal{E}=\left\{E=\left(
                         \begin{array}{cc}
                           .5 & .2i \\
                           0 & .5 \\
                         \end{array}
                       \right)\right\}\nonumber
                       \end{split}
\end{equation}
where $i=\sqrt{-1}$. It is not hard to see that $MN=NM$. A simple calculation leads to
\begin{equation}
\begin{split}
 wp(\mathcal{E})(M)=&\left(
                         \begin{array}{cc}
                           .5 & .2i \\
                           0 & .5 \\
                         \end{array}
                       \right)\left(
      \begin{array}{cc}
        .2 & 0 \\
        0 & 0 \\
      \end{array}
    \right)\left(
                         \begin{array}{cc}
                           .5 & .2i \\
                           0 & .5 \\
                         \end{array}
                       \right)^{\dagger}=\left(
                                           \begin{array}{cc}
                                             .05 & 0 \\
                                             0 & 0 \\
                                           \end{array}
                                         \right)
                       \nonumber\\
 wp(\mathcal{E})(N)=&\left(
                         \begin{array}{cc}
                           .5 & .2i \\
                           0 & .5 \\
                         \end{array}
                       \right)\left(
      \begin{array}{cc}
        .3 & 0 \\
        0 & .7 \\
      \end{array}
    \right)\left(
                         \begin{array}{cc}
                           .5 & .2i \\
                           0 & .5 \\
                         \end{array}
                       \right)^{\dagger}=\left(
                                           \begin{array}{cc}
                                             .103 & .07i \\
                                             -.07i & .175 \\
                                           \end{array}
                                         \right).
    \end{split}
\end{equation}
Thus,
\begin{equation}
\begin{split}
 wp(\mathcal{E})(M)\cdot wp(\mathcal{E})(N)=&\left(
                                                \begin{array}{cc}
                                                  .05 & 0 \\
                                                  0 & 0 \\
                                                \end{array}
                                              \right)
 \left(
   \begin{array}{cc}
     .103 & .07i \\
     -.07i & .175 \\
   \end{array}
 \right)=\left(
           \begin{array}{cc}
             .00515 & .0035i \\
             0 & 0 \\
           \end{array}
         \right)\nonumber\\
 wp(\mathcal{E})(N)\cdot  wp(\mathcal{E})(M)=& \left(
   \begin{array}{cc}
     .103 & .07i \\
     -.07i & .175 \\
   \end{array}
 \right)\left(
                                                \begin{array}{cc}
                                                  .05 & 0 \\
                                                  0 & 0 \\
                                                \end{array}
                                              \right)=
 \left(
   \begin{array}{cc}
     .00515 & 0 \\
     -.0035i & 0 \\
   \end{array}
 \right)
 \end{split}
\end{equation}
It is obvious that \[wp(\mathcal{E})(M)\cdot wp(\mathcal{E})(N)\neq wp(\mathcal{E})(N)\cdot wp(\mathcal{E})(M).\]

\end{exam}

\par
By virtue of Example \ref{exam2}, we immediately have the following
\begin{pro}\label{pro8}
 For any $n\geq 2$, there exist $M,N\in\mathcal{P}(\mathcal{H}_n)$ with $\mathrm{ dim}\mathcal{H}_n$=n, and $\mathcal{E}\in\mathcal{C}\mathcal{P}(\mathcal{H}_n)$ such that $MN=NM$ with
  \[wp(\mathcal{E})(M)\cdot wp(\mathcal{E})(N)\neq wp(\mathcal{E})(N)\cdot wp(\mathcal{E})(M).\quad\mbox{\qed}\]
\end{pro}

\section{Conclusions}\label{sec4}

In this paper, we have given some simple characterizations for the commutativity of quantum weakest preconditions. Further, we show by Example \ref{exam1} and Example \ref{exam2} that it is very difficult to characterize commutativity of quantum weakest preconditions in terms of $[M,N]$, which also can be seen from Proposition \ref{pro7} and Proposition \ref{pro8}.

\end{document}